\documentstyle[aps,multicol,epsfig]{revtex}
\begin{document}
\title{A Model for Granular Texture with Steric Exclusion}    
\author{H. Troadec$^1$, F. Radjai$^1$, S. Roux$^2$, and J.C. Charmet$^3$} 
\address{$^1$ LMGC, CNRS-Universit\'e Montpellier II, 
\\Place Eug\`ene Bataillon, 34095 Montpellier cedex, France. \\ 
$^2$ Laboratoire Surface du Verre et Interfaces, 
CNRS-Saint Gobain, \\39 Quai Lucien Lefranc,
93303 Aubervilliers Cedex, France. \\ 
$^3$ LPMMH, ESPCI, 10 rue Vauquelin, 75231 Paris Cedex 05, France.}
\maketitle

 \begin{abstract}
 We propose a new method to characterize the geometrical texture of a
granular packing  at the particle scale including the steric hindrance 
effect. This method is based on the assumption of a maximum  disorder
(entropy) compatible both with strain-induced anisotropy of the
contact network and steric exclusions.  We show that the predicted
statistics for the local configurations is in a fairly agreement
with our numerical data.    
\end{abstract}


\begin{multicols}{2}


 In modeling granular media from a microscopic point of view,    one is
naturally led to consider a fundamental question: how do the  properties
of the spatial organization of particles  control the behaviour at the
macroscopic scale?  A major difficulty is that the  effects arising from
the geometrical frustration of particles  prevail due to hard core
exclusions, and thus  the standard methods developped for multibody
systems fail as such  to provide a satisfactory description of many subtle
structural   features of a granular medium.

The most basic effect of mutual exclusions is that a
particle may be simultaneously touched by only a few neighbouring  
particles.  The coordination
number (the number of touching neighbours) 
cannot exceed 6 in a  2D assembly of particles of nearly the
same size and 12 in a 3D packing. 
These {\em local  environments} fluctuate strongly in space both in
terms of  the coordination number and the angular positions of the
neighbours.   

Another basic observation is that the relative angular
positions of the particles i.e. the directions of the contact
normals are not isotropically
distributed \cite{review,roux}.

\begin{figure}
\centerline{\psfig{figure=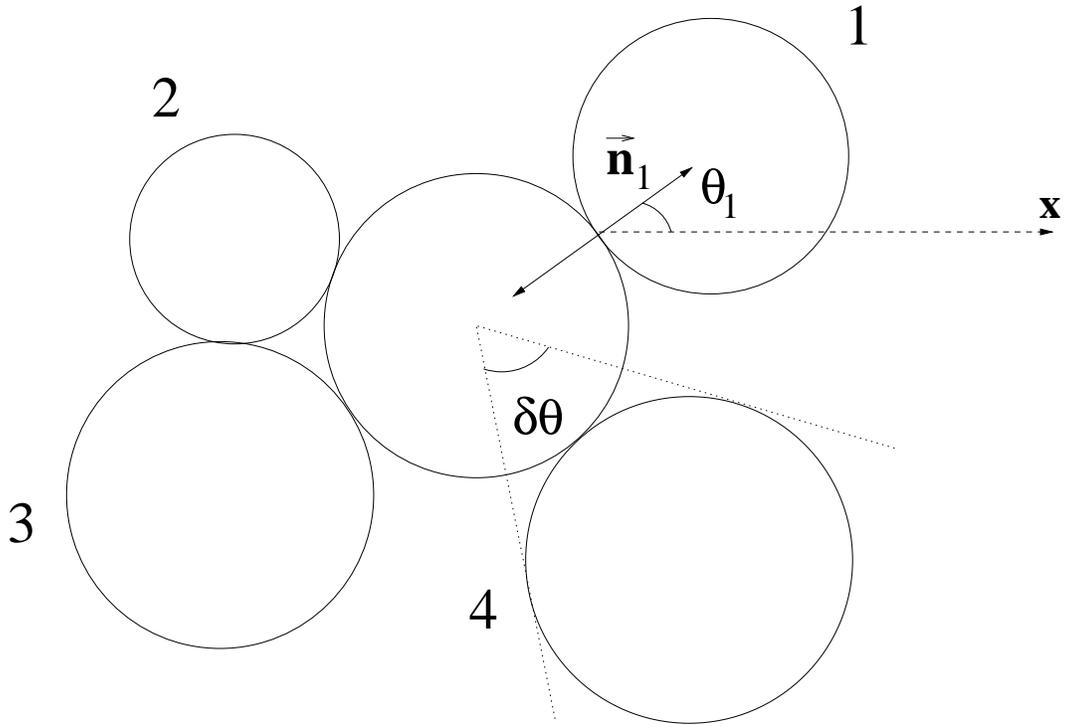,height=10.cm,angle=-0}}
\caption{Schematic representation of a particle 
environment with a coordination number $ z = 4 $. $ \theta_i $ is the angular position of particle $ i $ and $ \delta \theta $ is the exclusion angle \label{fig1}}        
\end{figure}

The adequate description of local environments including these 
geometrical features, i.e. a combination of excluded-volume 
constraints, disorder and anisotropy, 
is a crucial step in tracing back
the failure and flow properties of granular materials to 
the particle scale. 
Consider, for instance, a granular packing
in statistical equilibrium. The force balance on a particle
involves the angular positions of the few neighbours belonging
to the local environment. 
The fact that two neighbours may not
occupy two angular positions separated by an angle
smaller than a finite angle $ \delta \theta $
(approximately $ \frac{\pi}{3} $ for particles of nearly the same size,
see figure \ref{fig1})
reduces drastically the accessible equilibrium states of the 
particle and hence that of the whole packing. This effect
is even more critical at incipient failure and for larger anisotropies
of the contact directions. While the texture anisotropy has been
considered by several authors in the past \cite{review}, the key
role of steric exclusions has been almost completely
disregarded in micro-macro models of the quasi-static rheology
of granular materials.

In this paper, we propose a model that allows to construct local 
environments incorporating both the steric exclusions and  generic
disorder of granular media from   the global geometrical texture
described in terms of the  directions of interparticular contacts or the
corresponding anisotropy.   The basic assumption of our model is   a
{\em maximum directional disorder} consistent both with the texture  
anisotropy and with steric constraints. We apply our  approach to a
two-dimensional medium and solve the resulting  equation. The predicted
statistics of local environments  will then be compared with raw data
from numerical simulations.  


Let us consider the simple model of a 2D  granular assembly of rigid
disks.   In simulations or experiments in 2D geometry, a weak
polydispersity is necessary to avoid  local crystalline order. In our
theoretical description, as we shall see below, the geometrical 
disorder is explicitly taken into  account through the coordination
number and the contact directions. We may thus assume that particles are
of the same size,  and the steric hindrance   will be characterized by a
single exclusion  angle $ \delta \theta = \pi/3$.

The directional organization of the contact network is described by the
probability density of contact  normals
$p(\theta)$\cite{oda,cambou,bathurst,lanier}. This is  the probability
that a given particle has a contact along the direction  parametrized by
$\theta$, the polar angle of the contact  normal $\vec n$;
figure~\ref{fig1}. This distribution is induced by the relative motions
of particles, and so it  essentially reflects the deformation history or
the dynamics of the preparation process.  The function $p(\theta)$ is
$\pi$-periodic. It is often assumed that it can be   approximated by its
truncated Fourier transform \cite{bathurst}
  \begin{equation} 
  p(\theta) = \frac{1}{\pi} 
  \left\{1 + a \cos 2 (\theta - \theta_p) 
  \right\}, 
  \label{eqn1}
  \end{equation}
 where $a$ represents the anisotropy of the texture and $\theta_p $ is 
the average contact direction.  This functional form is reasonable for a
simple deformation history, but it may  considerably deviate from this
simple form otherwise \cite{cambou,lanier}. In order to bypass a
particular fitting form such as equation \ref{eqn1}  for characterizing
the anisotropy, one  may more generally construct  the ``fabric tensor''
$ { \mbox{\boldmath $ F $}} \equiv \langle \vec n \otimes \vec n\rangle$
where the  brackets denote averaging over all contacts in a
representative element of volume, and $\otimes$ is the dyadic
product\cite{satake,goddard}.  Then, the texture anisotropy $ a $ is
defined to be $ a = 2 | F_1 - F_2 |  $ where $ F_1 $ and $ F_2 $ are the
eigenvalues of $  { \mbox{\boldmath $ F $}} $.

A richer description of the texture is provided  by the probability
density $ f_z(\theta) $ that a particle with $z$  contacts has one
contact along $\theta$.    This function is defined over the range $[ 0,
2 \pi]$. Like $p(\theta)$,  it can be Fourier expanded, but the
corresponding anisotropy coefficients  now will depend on $z$. At 
leading order in $\theta$, we have
  \begin{equation} 
  f_z(\theta) = \frac{1}{2 \pi} 
  \left\{1 + a_z \cos 2 (\theta - \theta_z) 
  \right\}.
  \label{eqn1-1}
  \end{equation}

It is important to point out here that the steric  exclusions do not
allow for arbitrary global distributions $ f_z $. For $ f_z $ defined by
Equation \ref{eqn1-1}, the anisotropy $ a_z $  could take any values in
the range $[0,1]$  ($a_z \geq 1$ implies negative  values of $f_z$ for
some directions, whereas negative values can be avoided by  turning
$\theta_z$ into $\theta_z+\pi/2$). Nevertheless, the steric exclusions
impose an upper bound $a_{max}$  on the anisotropy. Indeed,  at most one
contact can be found within an angular sector of $\pi / 3$  around a
particle, i.e. 
  \begin{equation}
  \int_{\theta-\frac{\pi}{6}}^{\theta+\frac{\pi}{6}} 
  z f_z(\theta) d \theta \le 1 
  \label{eqn10}
  \end{equation}       
which together with the expression of $ f_z $ given Eq.~(\ref{eqn1-1}) yields 
  \begin{equation}
  a_{max}(z) = \frac{4 \pi}{\sqrt{3}} \left(\frac{1}{z}-\frac{1}{6} \right).
  \label{eqn11}
  \end{equation}
 This means that $ a_{max} $ decreases with $z$ and becomes zero for
$z=6$. In order to evaluate $f_z$ for a granular sample, the subset of 
particles with $z$ contacts are to be singled out.  Our numerical
simulations show that $a_z$ indeed decreases with $z$\cite{troadec}.


None of the above functions however directly provides an exhaustive
representation of the local environment of a particle.  
A complete description 
requires the coordination number $z$ of the particle and the directions 
$\theta_1, \ldots , \theta_z$ of the  contact normals around it.  Those
angles should fulfill the constraint of steric exclusion around
particles, so that an interval of width $\pi/3$ centered on the normal
direction of any contact is excluded.   In this way, the  particle
environments should be described through {\em multicontact} probability
distributions   $g_z(\theta_1, \cdots, \theta_z)$.  This distribution is
$2\pi$-periodic and its integration   over all contact directions but
one,   $ \theta_k $, gives back the global distribution   
$f_z(\theta_k) $ defined above: 
  \begin{equation}
  \int_{{\cal E}_z(k)} g_z(\{\theta_i\}) d\{\theta_i\}_{i \neq k}
  = f_z(\theta_k),
  \label{eqn2}
  \end{equation}
  where $ {\cal E}_z(k) $ is the domain corresponding  to integration
from $ 0 $ to $ 2 \pi $ over all the  $ \theta_i$ except $\theta_k$. 


Here, we would like    to construct the local distributions
$g_z(\{\theta_i\})$ from the  global distribution $f_z(\theta)$. Of
course, the solution is not unique  since $g_z$ contains a much richer
information  than $f_z$. The point, however, is to get a solution that
incorportes the required local information  with no bias towards a
particular solution.     In the absence of local steric constraints,
the most {\em unbiased}   situation implies that
  \begin{equation}
  g_z(\theta_1, \cdots,\theta_z)
  = \prod_{i=1}^z f_z(\theta_i).
  \label{eqn3}
  \end{equation}
  In our case, this solution is wrong since the steric  exclusions
require
  \begin{equation}
  g_z(\{\theta_i\})= 0 \ \ \ {\rm if} \ \ |\theta_i-\theta_j|<\pi/3,
  \label{eqn4}
  \end{equation}
  for $ \ i \neq j $ which is obviously not satisfied  by the above
solution. 

The solution that we propose is still to resort to a similar least
biased estimate, provided the above steric constraints are taken into
account.  Operationnally, this translates into the maximization of an
entropy functional $S [g_z]$ under constraints. According to Shannon's
entropy, we have
  \begin{equation}
  S [g_z]=  - \int_{{\cal D}_z} g_z(\{\theta_i\}) \log[g_z(\{\theta_i\})] 
  d\{\theta_i\},
  \label{eqn5}
  \end{equation}
  where ${\cal D}_z = ([0,2\pi])^z$ is the integration domain.  This
functional should be maximized over the set  of functions that  fulfill
both the steric exclusions, Eq.~(\ref{eqn4}), and the normalization
conditions,  Eq.~(\ref{eqn2}).  We note that the entropy formalism, as a
basic statistical tool, has been previously applied to granular
materials in  order to characterize other aspects of a random packing
such as the coordination number and the void ratio \cite{backman}.


Equation~(\ref{eqn4}) implies that  the maximization of $S$ has to be
restricted to the region   ${\cal A}_z$ of ${\cal D}_z$ that is allowed
by steric   exclusions. Indeed, the maximization applies only to the 
lacking information, and here the value of $g_z$ is already  known
($g_z=0$) for the angles belonging to the excluded  region ${\cal B}_z
={\cal D}_z - {\cal A}_z$.  Hence, the above restriction allows to
account for the steric constraints.   

On the other hand, the normalization  conditions~(Eq \ref{eqn2}.) can be
imposed through Lagrange  multipliers. This leads to the maximization of
the functional 
  \begin{equation}
  T [g_z] = S[g_z] - \sum_{i=1}^z  \int_0^{2\pi}  
  \lambda_i(\theta_i)  C_i  [g_z] \  d\theta_i
  \label{eqn6}
  \end{equation}
  over ${\cal A}_z$, where the $\lambda_i(\theta_i)$ are the Lagrange 
multipliers and  
  \begin{equation}
  C_i  [g_z] =  f_z(\theta_i) - \int_{{\cal E}_z(i)} g_z(\{\theta_i\})
  d\{\theta_k\}_{k\neq i}. 
  \label{eqn7}
  \end{equation}


The solution is determined by setting the functional derivative  of $T 
[g_z]$ to zero.  Let us remark that here the only correlations among the
angles are the steric exclusions themselves. This means that   the
angles $ \theta_i $ can be considered as independant  variables when
they are restricted to the allowed  domain $ A_z $.  As a consequence,
the solution is given by a product of identical functions  $ h $ of a
single angle in the allowed domain. In this way, the general solution is
written as follows
  \begin{equation}
  g_z(\{\theta_i\})= \left( \prod_{i\neq j} G(\theta_i-\theta_j)\right)
  \prod_{k=1}^z h_z(\theta_k), 
  \label{eqn8}
  \end{equation}
 where $G(\theta)$ is a $2\pi$-periodic function such that $G(\theta)=0$
for $|\theta|<\pi/3$, and $G=1$ otherwise.  These prefactors take care
of the steric exclusions.  Note that without these exclusions, we have $
G = 1 $ for all angles  and the  solution~(\ref{eqn3})  is recovered
with $h_z(\theta) \equiv f_z(\theta)$.

The function $h_z$ is determined by the normalization 
condition~(\ref{eqn2}) which becomes now the  following implicit
equation in terms of $h$:   
  \begin{equation}
  \int_{{\cal A}_z(i)} \prod_{k=1}^z h_z(\theta_k)  
  d\{\theta_j\}_{j\neq i}= f_z(\theta_i), 
  \label{eqn9} 
  \end{equation}
  where $ {\cal A}_z(i) = {\cal E}_z(i) \cap {\cal A}_z$. This equation
can be used for a numerical computation of  $ h_z $ for a given $ f_z $.
We used a simple iterative  scheme over the functions $u_n(\theta)$
satisfying
  \begin{equation} 
  u_{n+1} (\theta_1) = \epsilon u_n + (1 - \epsilon)
  \frac{f_z(\theta_1)}{ \int_{A_z(1)} u_n (\theta_2) \cdots u_n(\theta_z)
  d\theta_2 \cdots d \theta_z}.
  \end{equation}
  where $ \epsilon $ is a relaxation parameter.

\begin{figure}
\centerline{\psfig{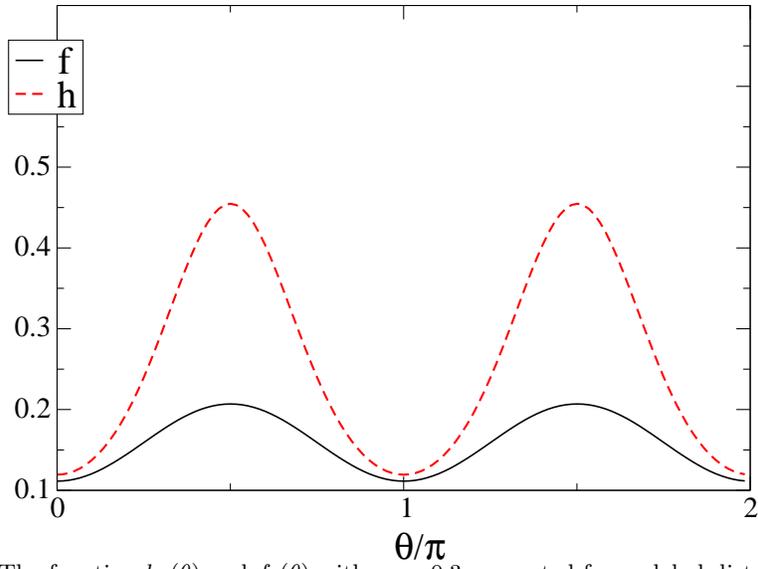}}
\caption{The function $h_3(\theta)$ and $f_3(\theta)$ with $a_3=0.3$ computed for 
a global distribution. \label{fig2}}        
\end{figure}

The solution $h_z$ of  equation~(\ref{eqn9}) is simply a   fixed point
of the latter recursion.  Our iterative process converges smoothly to the
solution when the latter exists and, moreover, its convergence
domain matches exactly the set of admissibles values of the anisotropy
according to equation (\ref{eqn11}).  We show one example of solution in
Figure~\ref{fig2}: The function  $h_3(\theta)$ was computed according to
the above procedure with  $f_3(\theta)$ given by Eq.~(\ref{eqn1-1}) for 
$z=3$ and $a_3=0.3$. We see that $h_3$ has the same monotony as $f_3$,
but it  deviates from a simple sinusoidal form.


Of course, it is desirable to compare the multicontact 
distributions $g_z$ predicted 
by the above approach with direct data from experimental or 
numerical observations. However, for a sufficient 
statistical precision requires very large samples. 
In fact, given a granular sample, the subset of particles 
with $z$ contacts are considered. Assuming that in a 2D 
system of disks, the number of particles 
is the same for each of the prevailing values 3, 4 and 5 of $ z $, 
which is actually not the case since there are 
typically more particles with four contacts, 
then only nearly one third of the particles 
is available for the evaluation of $g_z$. On the other hand, subdividing the 
interval $[0,2\pi]$ into $n$ angular sectors and requiring on average 
$m$ events in each elementary box in $[0,2\pi]^z$, one would need a sample of 
$m (n^z)$ contacts, or approximately $(2m/z) n^z$ particles. 
For $z=4$, $n=20$ and $m=100$, this amounts to $8.10^6$ particles, which 
is practically inaccessible with the present-time computation power. 

\begin{figure}
\centerline{\psfig{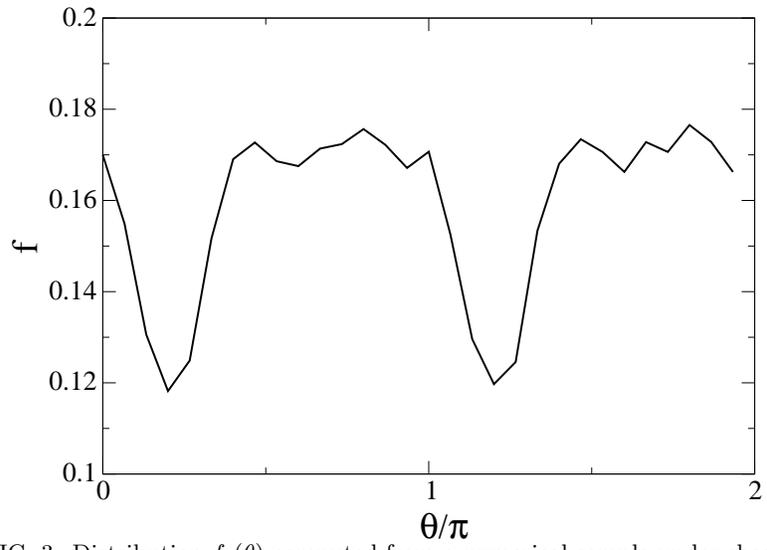}}
\caption{Distribution $f_4 (\theta)$ computed from a numerical sample 
under shear.\label{fig3}} 
\end{figure}

Nervertheless, we still may get a reliable comparison by focussing  on
the distributions  $p_d( \Delta \theta )$ of the difference  $ 
\Delta \theta = |\theta_1 - \theta_2| $ between contact directions 
extracted
from pair distributions $p_2(\theta_1,\theta_2)$. The latter was
calculated from numerical samples and  from the theoretical estimation
of $g_z$. To do so, we performed  numerical simulations of a system  of
4000 particles with bi-periodic boundary conditions by means of the
molecular dynamics method.  We used a viscous-regularized Coulomb
friction law and a linear  spring-dashpot model for particle
interactions \cite{Luding}.  The coefficient of friction was $0.5$ and
the particle radii   were uniformly distributed between $R_1$ and $R_2$
with $R_2= 2 R_1$.  The system  was subjected to simple shearing and
thanks to bi-periodic  boundary conditions we were able to keep shearing
in the steady state  for very large strains without any wall effect. 
Figure~\ref{fig3} shows the function $f_4(\theta)$
corresponding to the subset of  particles  with four contacts.          
The global steady-state anisotropy, $a_z$, calculated from the fabric
tensor was about $0.13$.
   
The distribution $p_d$ was calculated over the cumulated data from 
several well-separated snapshots of the steady state for the set of
particles with coordination number $ z $. For the  theoretical
evaluation of $p_d$ corresponding to these particles,  we used the
numerical distributions $f_z(\theta)$ for the calcultation of  $g_z$
following our approach, from which the distributions $ p_2 $ and $
p_d $ were extracted. 

\begin{figure}
\centerline{\psfig{figure=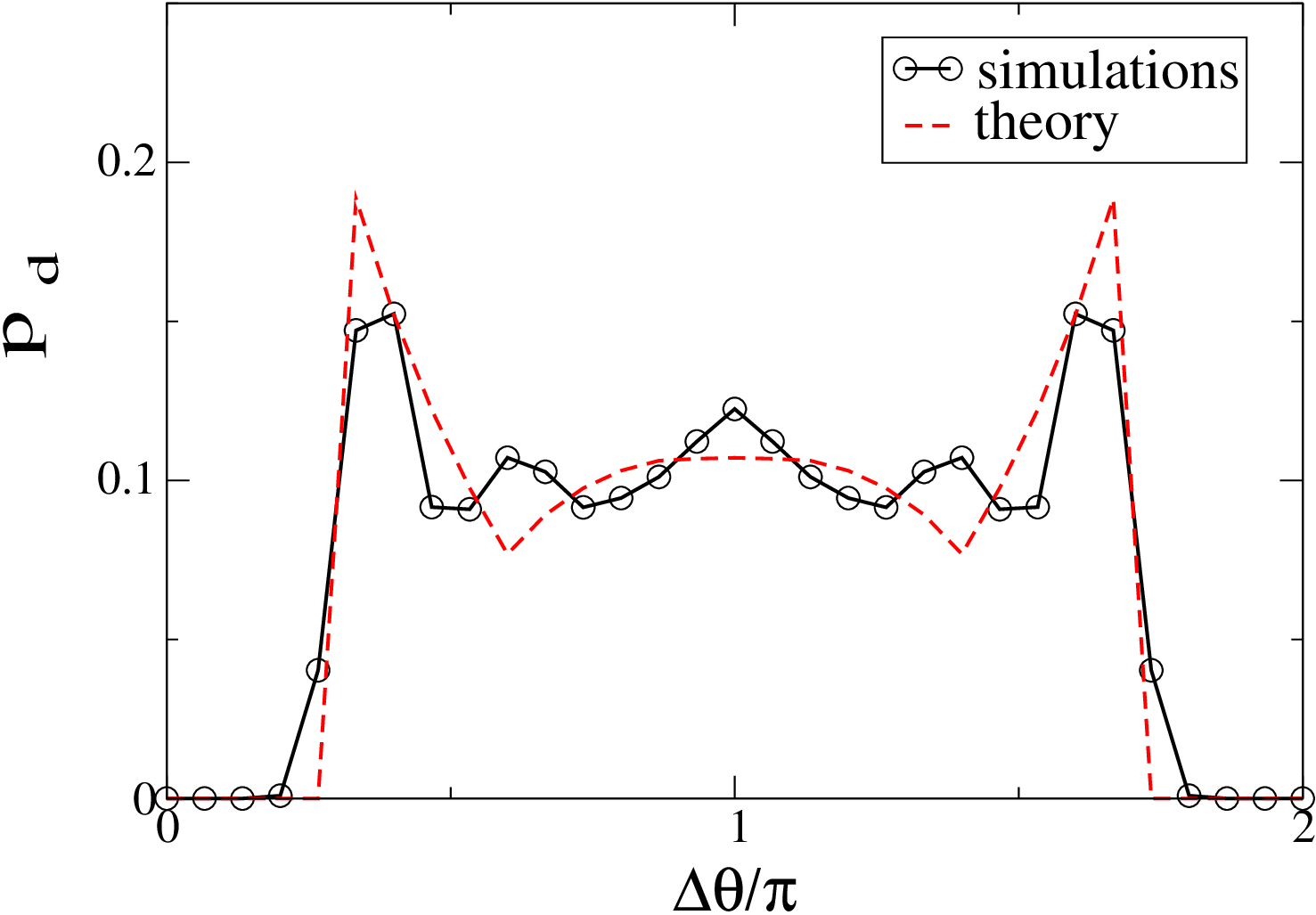,height=10.cm,angle=-90}}
\caption{Distribution $p(\Delta \theta)$ of the difference between 
successive contacts of a particle with four contacts as predicted by 
our model and obtained from numerical simulations.\label{fig4}}      
\end{figure}

\begin{figure}
\centerline{\psfig{figure=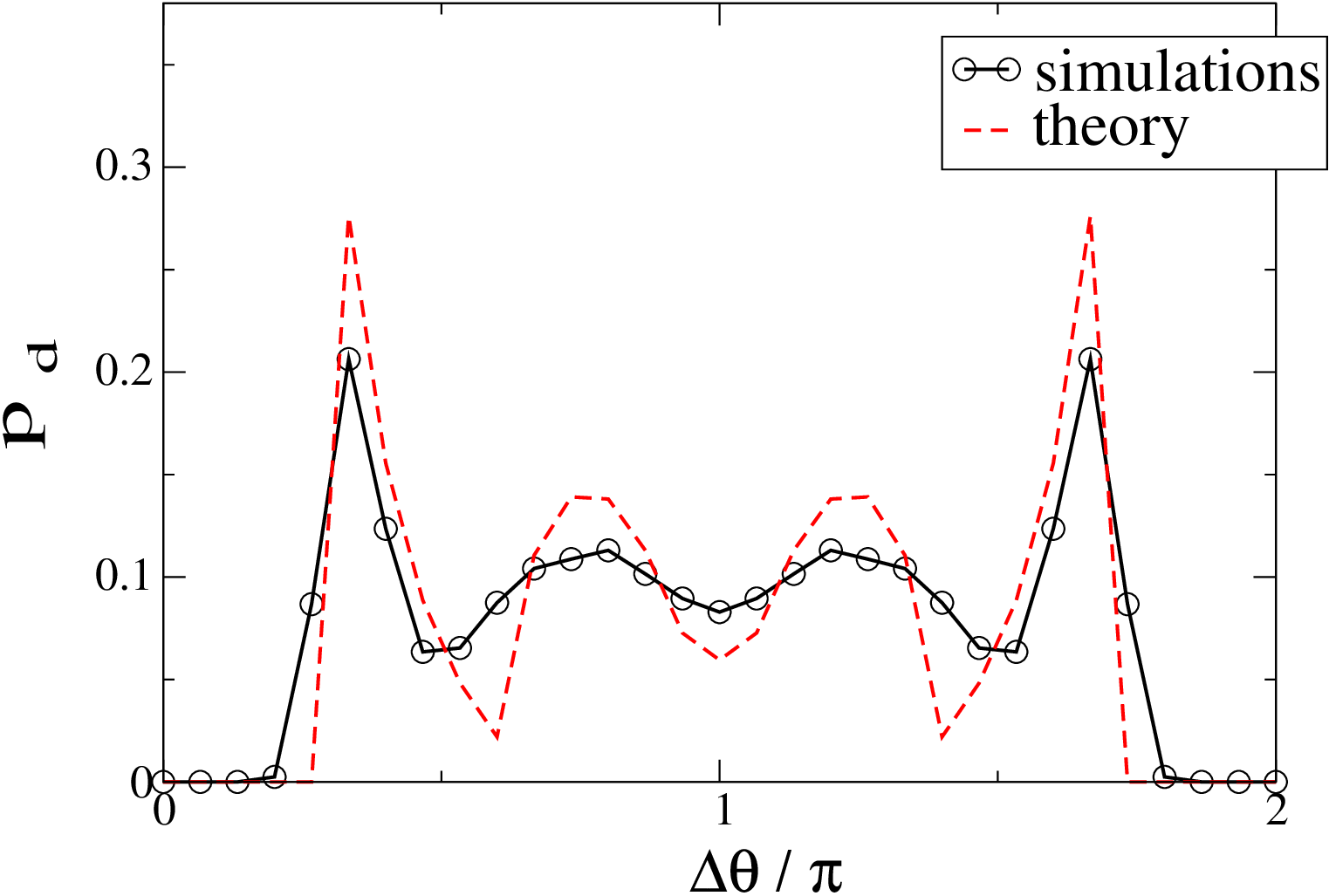,height=10.cm,angle=-90}}
\caption{Distribution $p(\Delta \theta)$ of the difference between 
successive contacts of a particle with five contacts as predicted by 
our model and obtained from numerical simulations.\label{fig5}}      
\end{figure}

In figures  \ref{fig4} and \ref{fig5}, we have plotted  theoretical and
numerical $ p_d $  as a function of $ \Delta \theta $ for $ z = 4 $
and $ z = 5 $,  respectively.  Although only the geometrical constraints
are taken into account for the theoretical construction of $ g_z $ (no
force considerations),  we see that the predicted distributions $
p_d $ both for $ z = 4 $ and $ z = 5 $ fit fairly well the
numerical data. The fit is nevertheless better for $ z = 5 $ than for $
z = 4 $. On the other hand, the fit for $ z = 3 $ (not shown) is much
less satisfactory. In fact, in going from $ z = 3 $ to $ z = 5 $, the
requirement of force balance becomes less and less stringent, whereas
steric exclusions dominate more and more the behaviour. Let us also
remark that  for $ z = 4 $ a small peak appears on the  numerical curve 
at $\Delta \theta = 2\pi / 3 $, as can be seen in  figure \ref{fig4}.
This peak, which is absent from the theoretical curve, is probably the
signature of a local order induced by the  rather small polydispersity
of our sample. 


These results suggest that, if the steric exclusions are correctly taken
into account, a satisfactory evaluation of  the statistics of local
environments in terms of the functions $ g_z $ can be obtained at least
for $ z = 4 $ and $ z = 5 $.  The steric exclusions control  to a large
extent the local distributions  (figures \ref{fig4}, \ref{fig5})  and
reduce the range of admissible anisotropies  (equation \ref{eqn11})
\cite{troadec}.

A natural extension of this model, at the cost of additional 
parameters, is  to include the force  balance as well as the steric
exclusions. This  would allow to tackle the case of $ z = 3 $ more
accurately  than we did in this paper.  The idea is to include both
contact directions and  contact forces in the model.  The forces are
coupled to a local stress tensor in the same  way as the contact 
directions were coupled to the fabric tensor.  Then, the entropy of
local distributions is maximized  regarding constraints arising from
steric exclusions, but also the force balance, and constitutive laws 
such as Coulomb friction and Signorini's condition (positivity of 
normal forces).  

Apart from its importance in itself as a  general model for the granular
texture including steric exclusions, the characterization of the texture
in terms  of local environments is a crucial step in order to relate the
macroscopic behaviour to the particle scale. For example, the range of
admissible stresses  can be studied by considering representative 
configurations generated according to the multicontact statistics of
local  environments \cite{troadec}. A similar procedure based on entropy
formalism may be applied as well to other relevant 
microscopic  configurations
such as the cells composed of contiguous 
particles that carry the most local strains at the microscopic 
strains.

\end{multicols}

\clearpage

\clearpage

\clearpage

\clearpage

\clearpage

\end{document}